\documentclass{moriond}

\def\be{\begin{equation}}
\def\ee{\end{equation}}
\def\bea{\begin{eqnarray}}
\def\eea{\end{eqnarray}}

\usepackage{amsmath,amssymb}
\usepackage{mathrsfs}
\usepackage{amsfonts} 
\newcommand{\SUtwo}{\text{SU}(2)_L} %Grupo Simetria Su(2)
\newcommand{\refeq}[1]{eq.\,\eqref{#1}} % reference to equation ``eq. (n)''
\newcommand{\re}[1]{\text{Re}\left(#1\right)}

\newcommand{\Hc}{\text{h.c.}}
\newcommand{\Zn}[1]{\mathbb{Z}_{#1}}
\newcommand{\ZZ}{\Zn{2}}
\newcommand{\tb}{t_\beta}

\newcommand{\tbinv}{\tb^{-1}}

\newcommand{\nh}{\mathrm{h}}

\newcommand{\nH}{\mathrm{H}}
\newcommand{\nA}{\mathrm{A}}
\newcommand{\nS}{\mathrm{S}}
\newcommand{\cH}{\mathrm{H}^\pm}

\newcommand{\mH}{m_{\nH}}
\newcommand{\mA}{m_{\nA}}
\newcommand{\mcH}{m_{\cH}}
\newcommand{\mS}[1]{m_{\mathrm{S}_{#1}}}
\newcommand{\nl}[1]{n_{#1}}

\newcommand{\nrl}[1]{\re{n_{#1}}}
\newcommand{\nrle}{\nrl{e}}
\newcommand{\nrlm}{\nrl{\mu}}
\newcommand{\nrlt}{\nrl{\tau}}

\newcommand{\FQ}{Q}
\newcommand{\FL}{L}
\newcommand{\Fu}{u}
\newcommand{\Fd}{d}
\newcommand{\Fl}{\ell}

\newcommand{\ferX}[3]{{#1}_{#2#3}}
\newcommand{\ferXb}[3]{\bar #1_{#2#3}}

\newcommand{\dR}[1]{\ferX{\Fd}{R}{#1}}

\newcommand{\uR}[1]{\ferX{\Fu}{R}{#1}}

\newcommand{\lR}[1]{\ferX{\Fl}{R}{#1}}

\newcommand{\Hv}{H_1}
\newcommand{\Hvd}{H_1^\dagger}
\newcommand{\Hvti}{\tilde H_1}
\newcommand{\Ho}{H_2}
\newcommand{\Hod}{H_2^\dagger}
\newcommand{\Hoti}{\tilde H_2}

\newcommand{\QLb}[1]{\ferXb{\FQ}{L}{#1}}

\newcommand{\LLb}[1]{\ferXb{\FL}{L}{#1}}

\newcommand{\matXF}[2]{{\rm #1}_{#2}}

\newcommand{\basematM}{M}
\newcommand{\basematN}{N}
\newcommand{\matNf}[1]{\matXF{\basematN}{#1}}

\newcommand{\matMf}[1]{\matXF{\basematM}{#1}}

\newcommand{\matND}{\matXF{\basematN}{d}}

\newcommand{\matNU}{\matXF{\basematN}{u}}

\newcommand{\matNL}{\matXF{\basematN}{\ell}}

\newcommand{\matMD}{\matXF{\basematM}{d}}

\newcommand{\matMU}{\matXF{\basematM}{u}}

\newcommand{\matML}{\matXF{\basematM}{\ell}}

\usepackage{bm}
\newcommand{\Photo}{}

\begin{document}
\vspace*{4cm}
\title{LEPTONIC $\bm{g-2}$ IN 2HDM}

\author{ FRANCISCO J. BOTELLA }
\author{ FERNANDO CORNET-GOMEZ$^{}$\,\footnote{Speaker and corresponding author: Fernando.Cornet@ific.uv.es} }
\author{ CARLOS MIRÓ }
\author{ MIGUEL NEBOT }

\address{Departament de Física Teòrica and IFIC, Universitat de València-CSIC\\
Instituto de Física Corpuscular, C/Catedrático José Beltrán, 2. E-46980 Paterna, Spain.}

\maketitle\abstracts{The experimental observations of the electron and muon anomalous magnetic moment present discrepancies with respect to the Standard Model predictions. A class of flavor conserving Two Higgs Doublet model, stable under renormalization, that is capable of explaining both anomalies simultaneously is presented. This model can also explain an excess observed by ATLAS in $\sigma(pp\to\nS)_{[\text{ggF}]}\times \text{Br}(\nS\to\tau^+\tau^-)$. }

\section{Introduction}
In this proceeding, we review and extend the works presented in Refs.\cite{BotellaOlcina:2020kvb,Botella:2020xzf,Botella:2022rte} regarding the simultaneous explanation of the electron and muon $g-2$ anomalies. Furthermore, we present a preliminary result concerning an excess in the di-tau channel of a heavy Higgs with a mass of  $m_S\sim400$ GeV observed by ATLAS \cite{ATLAS:2020zms}. The most recent determination of the anomalous magnetic moment of the muon \cite{Muong-2:2021ojo,Muong-2:2021vma} enforces the long-standing discrepancy (4.2$\sigma$) between the Standard Model (SM) prediction \cite{Aoyama:2020ynm} and the experimental observation
\begin{equation}\label{eq:damu}
 \delta a_\mu^{\rm Exp}=a_\mu^{\rm Exp}-a_\mu^{\rm SM}=(2.5\pm 0.6)\times 10^{-9}\,.
\end{equation}
There is a recent lattice calculation \cite{Borsanyi:2020mff} of the Hadronic Vacuum Polarization (HVP) contribution that shifts the SM prediction to the experimental value, solving the existing discrepancy. However, it was shown in Ref. \cite{Crivellin:2020zul} that solving the discrepancy with the HVP contribution would create a tension of the same size in the electroweak precision data fits. Until this lattice calculation is crosschecked by other collaborations, we assume the SM prediction of the $(g-2)_\mu$ as the one provided by the White Paper 2020 \cite{Aoyama:2020ynm}, which is in tension with the experimental observation.
 
Regarding the electron, the situation is not so clear. There are two different observations that are incompatible between them and that also deviate slightly from the SM prediction. The observation from atomic recoil using Cesium provides \cite{Parker:2018vye}
\begin{equation}\label{eq:dae:Cs}
 \delta a_e^{\rm Exp,Cs}=-(8.7\pm 3.6)\times 10^{-13}\, ,
\end{equation}
that imply a 2.4 $\sigma$ deviation, while the one using Rubidium \cite{Morel:2020dww}
\begin{equation}\label{eq:dae:Rb}
 \delta a_e^{\rm Exp,Rb}=(4.8\pm 3.0)\times 10^{-13}\, ,
\end{equation}
only deviates 1.6$\sigma$ but in the opposite direction. 

We will focus our attention on discussing $\delta a_e^{\rm Exp,Cs}$, since it is the hardest to explain with scalar mediators due to the different sign with respect to $\delta a_\mu^{\rm Exp}$, but we will also show some results of the $\delta a_e^{\rm Exp,Rb}$ analysis.

\begin{figure}
\begin{minipage}{0.45\linewidth}
\centerline{\includegraphics[width=\linewidth]{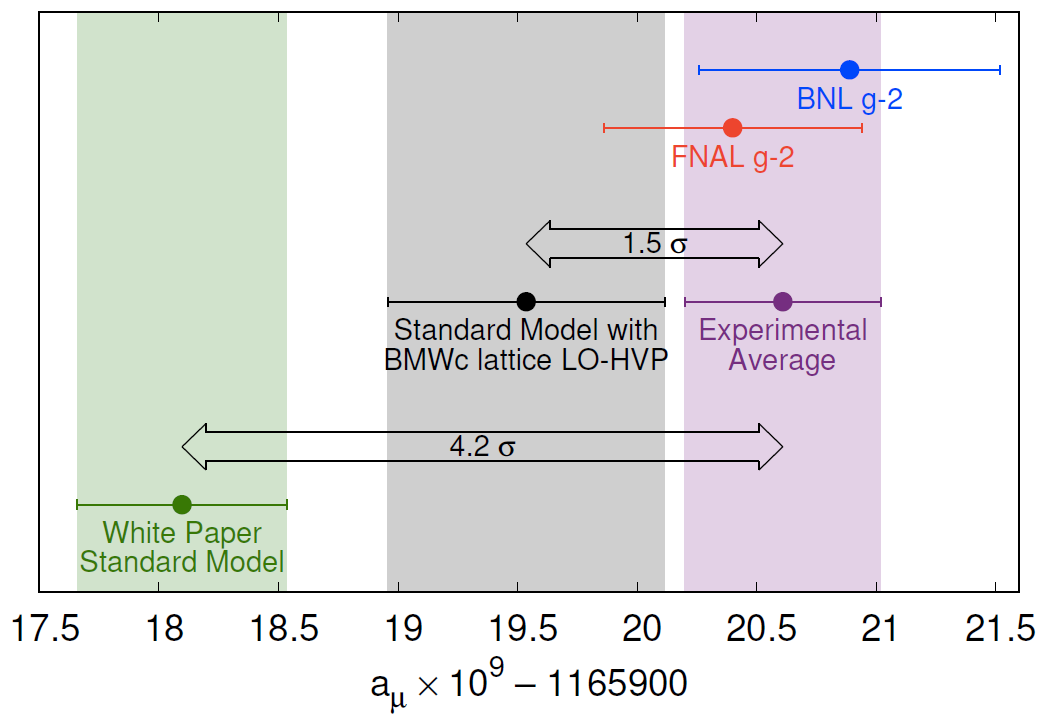}}
{\footnotesize a) Current situation with the experimental measurements and the different determinations of the muon $(g-2)$. Plot from L. Lellouch talk.}
\end{minipage}
\hfill
\begin{minipage}{0.45\linewidth}
\centerline{\includegraphics[width=0.7\linewidth]{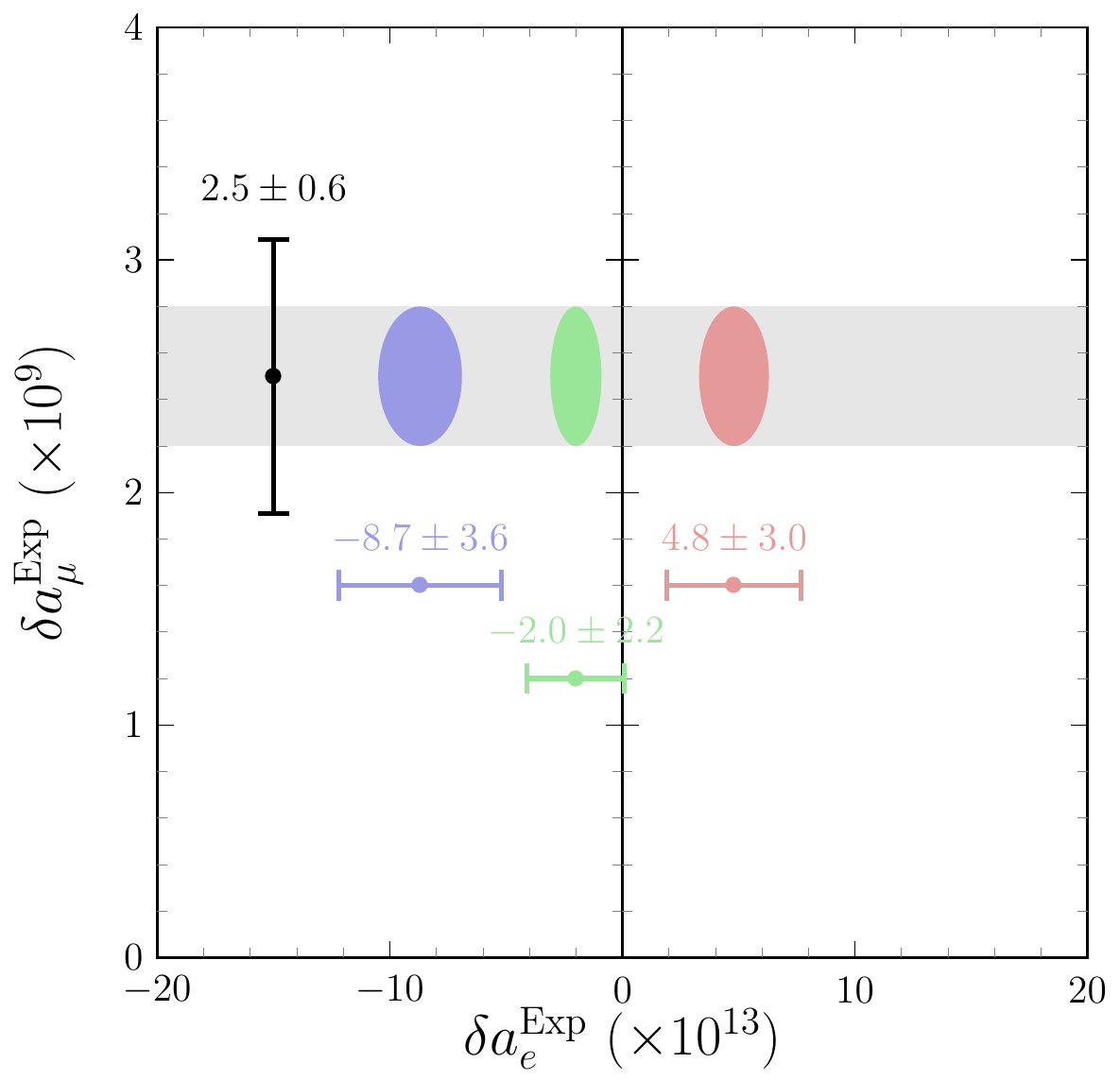}}
{\footnotesize b) Allowed $\delta a_\mu^{\rm Exp}$ vs. $\delta a_e^{\rm Exp}$ regions in the different analyses. The same color coding will be used in the following plots.} 
\end{minipage}
%\hfill
%\begin{minipage}{0.32\linewidth}
%\centerline{\includegraphics[width=\linewidth]{figures/w4_fig_02a.png}}
%\end{minipage}
\caption[]{Electron and muon anomalies studied in this proceedings.\label{Fig1}}
\label{fig:radish}
\end{figure}

\section{The model}
The 2HDM is a minimal extension of the SM where just a second scalar doublet of $\SUtwo$ is added. In this class of models, a new flavor structure, $\matNf{f}$, appears in the Yukawa sector:

\begin{equation}
\renewcommand{\arraystretch}{1.2}
\begin{array}{rc@{\,}c@{\,}l}
\mathscr L_{\rm Y}&=
 & -&\frac{\sqrt{2}}{v}\QLb{}\left(\Hv\matMD+\Ho\matND\right)\dR{}
 -\frac{\sqrt{2}}{v}\QLb{}\left(\Hvti\matMU+\Hoti\matNU\right)\uR{}\\
 & & -& \frac{\sqrt{2}}{v}\LLb{}\left(\Hv\matML+\Ho\matNL\right)\lR{}+\Hc \, ,\\
\end{array}
 \label{eq:YukLag}
\end{equation}
where $\matMf{f}$ are the diagonal fermion mass matrices. It is well-known that in the 2HDM, the matrices $\matNf{f}$ are a source of Flavor Changing Neutral Couplings (FCNC), since in general, they are not simultaneously diagonalizable with the mass matrices. Thus, the inclusion of a second Higgs doublet constitutes a challenge since FCNC are very constrained experimentally but it also opens the possibility of explaining observations that the SM cannot accommodate. To avoid FCNC we will apply general Flavor Conservation (gFC), that is, we impose that both Yukawa matrices are diagonalizable simultaneously. On the one hand, the quark sector of our model corresponds to a Type I 2HDM where the flavor matrices are proportional by the ratio of the two doublets vacuum expectation values, $\matNf{q}=\tbinv\matMf{q}$. On the other hand, the lepton matrix $\matNL$ is chosen to be arbitrary and diagonal (gFC)
\begin{equation}
\matNL= \begin{pmatrix}
n_e & 0 & 0\\ 0 & n_\mu & 0 \\ 0 & 0 & n_\tau
\end{pmatrix}\, ,
\end{equation}
 which is one-loop stable under Renormalization Group Evolution as was proved in Ref. \cite{Botella:2018gzy}. As the $g-2$ is a CP-conserving observable, we are not interested in including new sources of CP-violation thus the $n_\ell$ are defined as reals, and the scalar potential reads 
\begin{equation}
\renewcommand{\arraystretch}{1.2}
\begin{array}{rc@{\,}c@{\,}l}
V&=& &m_{11}^2\Hvd\Hv+m_{22}^2\Hod\Ho-\left(m_{12}^2\Hvd\Ho+\Hc\right)\nonumber\\
& &+&\frac{1}{2}\bar{\lambda}_1\left(\Hvd\Hv\right)^2+\frac{1}{2}\bar{\lambda}_2\left(\Hod\Ho\right)^2+\bar{\lambda}_3\left(\Hvd\Hv\right)\left(\Hod\Ho\right)+\bar{\lambda}_4\left(\Hvd\Ho\right)\left(\Hod\Hv\right)\, ,
\end{array}
 \label{eq:ScalarPot}
\end{equation}
where the term $\left(m_{12}^2\Hvd\Ho+\Hc\right)$ softly breaks the $\ZZ$ symmetry that generates the Type I in the quark sector. This soft-breaking allows us to reach scalar masses above 1 TeV and values of $\tb$ greater than 8 that are otherwise forbidden by perturbativity constraints. The fact that the scalar potential in \refeq{eq:ScalarPot} conserves CP leads to a scalar sector formed by two scalar fields, $\nh$, and $\nH$, one pseudoscalar $\nA$ and two charged particles $\cH$. In the so-called scalar alignment limit, the couplings of $\nh$ coincide with those of the SM Higgs ($\nh_{125}$). 

\section{Analysis}
The main goal of the work presented here is to explain both the electron and muon anomalies simultaneously. We can split the prediction for the anomalous magnetic moment, $a_{\ell}^{\rm Th}$, in the SM prediction, $a_{\ell}^{\rm SM}$, plus the new contribution coming from the 2HDM, $\delta a_{\ell}$, as
\begin{equation}
a_{\ell}^{\rm Th}=a_{\ell}^{\rm SM}+\delta a_{\ell},
\end{equation}
and we can parametrize the New Physics (NP) contribution as 
\begin{equation}
 \delta a_\ell=K_\ell\,\Delta_\ell,\qquad K_\ell=\frac{1}{8\pi^2}\left(\frac{m_\ell}{v}\right)^2.
\end{equation}
To reproduce the experimental observations we need to impose $\delta a_{\ell}=\delta a_{\ell}^{\text{Exp}}$ what fixes the values of $\Delta_\ell$ in the different scenarios to
\begin{equation}
\Delta_\mu\simeq 1, \qquad \Delta_e^{\rm Cs}\simeq -16,\qquad \Delta_e^{\rm Rb}\simeq 9\,.
\end{equation}
It is well-known that both the one and two loops contributions can be relevant. The one-loop contribution contains diagrams with neutral and charged mediators ($\nH$, $\nA$ and $\cH$) 
\begin{equation}\label{eq:deltal:1loop}
\Delta_{\ell}^{(1)}= |\nl{\ell}|^{2}\left(\frac{I_{\ell\nH}}{\mH^{2}}-\frac{I_{\ell\nA}-2/3}{\mA^{2}}-\frac{1}{6{\mcH}^{2}}\right)
\end{equation}
where the loop function reads
\begin{equation}
 I_{\ell S}\simeq-\frac{7}{6}-2\ln\left(\frac{m_\ell}{m_S}\right)\,.
\end{equation}
Among all the possible two-loop diagrams, the dominating ones are the Barr-Zee which contribute with
\begin{equation}\label{eq:deltal:2loop}
 \Delta_\ell^{(2)}=-\frac{2\alpha}{\pi}\,\frac{\nl{\ell}}{m_\ell}\,F\,,
\end{equation}
where $F$, a quantity that is independent of $\ell$, is defined as
\begin{equation}\label{eq:F:2loop}
 F=\frac{\tbinv}{3}\left[4(f_{t\nH}+g_{t\nA})+(f_{b\nH}-g_{b\nA})\right]
 +\frac{\nrlt}{m_\tau}(f_{\tau\nH}-g_{\tau\nA})+\frac{\nrlm}{m_\mu}(f_{\mu\nH}-g_{\mu\nA}).
\end{equation}
The two-loop functions $f(z)$ and $g(z)$ are defined as
\begin{equation}
\renewcommand{\arraystretch}{1.5}
\begin{array}{rc@{\,}c@{\,}l}
f(z)& = \frac{z}{2}\int_0^1 dx\,\frac{1-2x(1-x)}{x(1-x)-z}\,\ln\left(\frac{x(1-x)}{z}\right)\,,\quad
g(z)& = \frac{z}{2}\int_0^1 dx\,\frac{1}{x(1-x)-z}\,\ln\left(\frac{x(1-x)}{z}\right)\,.
\end{array}
\end{equation}

In addition to the $g-2$ anomalies, a big amount of available observables are included in the fit in order to constrain the parameter space:
\begin{itemize}
\item Scalar sector: the scalar potential is imposed to be bounded from below. Perturbative unitarity of $2\to 2$ scattering amplitudes is also required. As it was mentioned before, the  $\ZZ$ symmetry is broken by having $\mu_{12}^2\neq0$ to allow heavy scalars and $\tb>8$. Electroweak precision data is included through the oblique parameters.

\item The Yukawa couplings must remain perturbative:
\begin{equation}
\frac{|n_f|}{v}\leq\mathcal{O}(1)\quad \Rightarrow \quad |n_f|\lesssim 250\, \text{GeV}
\end{equation}

\item Signal strengths data constrain the prediction of cross section times branching ratio of the scalar $\nh$. The current data favors the alignment limit. 

\item Lepton flavor universality including leptonic and semileptonic decays.

\item Flavor constraints: neutral meson-mixing and $b\to s\gamma$.

\item LEP data, in particular $e^{+}e^{-}\to\ell^{+}\ell^{-}$.

\item LHC direct searches of extra scalar and pseudoscalar particles.
\end{itemize}
For further details on the predictions in this model of the different observables and the experimental values used in the fits see Refs. \cite{Botella:2020xzf,Botella:2022rte}.

\section{Results}

\begin{figure}[!htb]
\begin{center}
\includegraphics[width=0.32\textwidth]{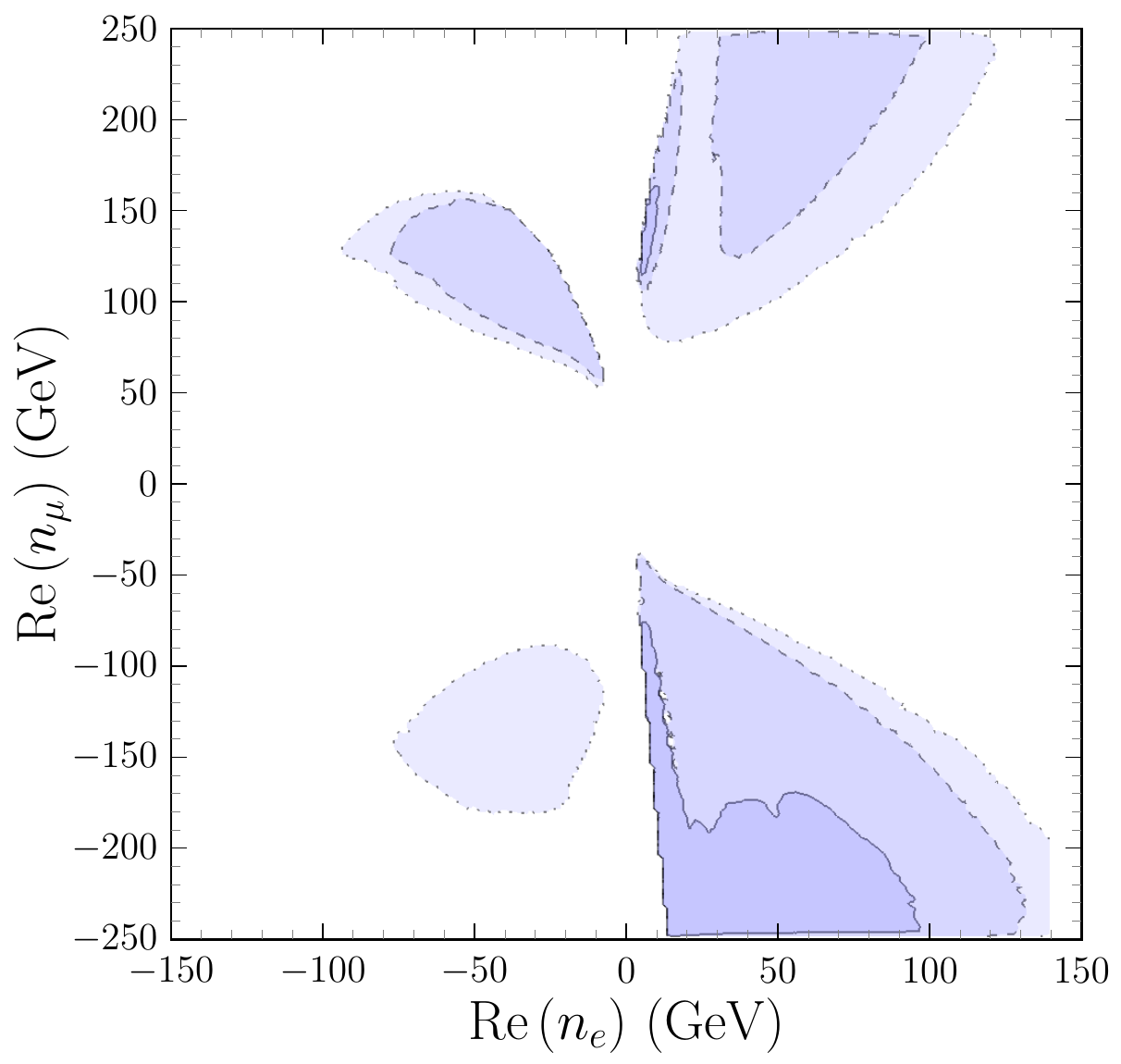}
\hfill
\includegraphics[width=0.32\textwidth]{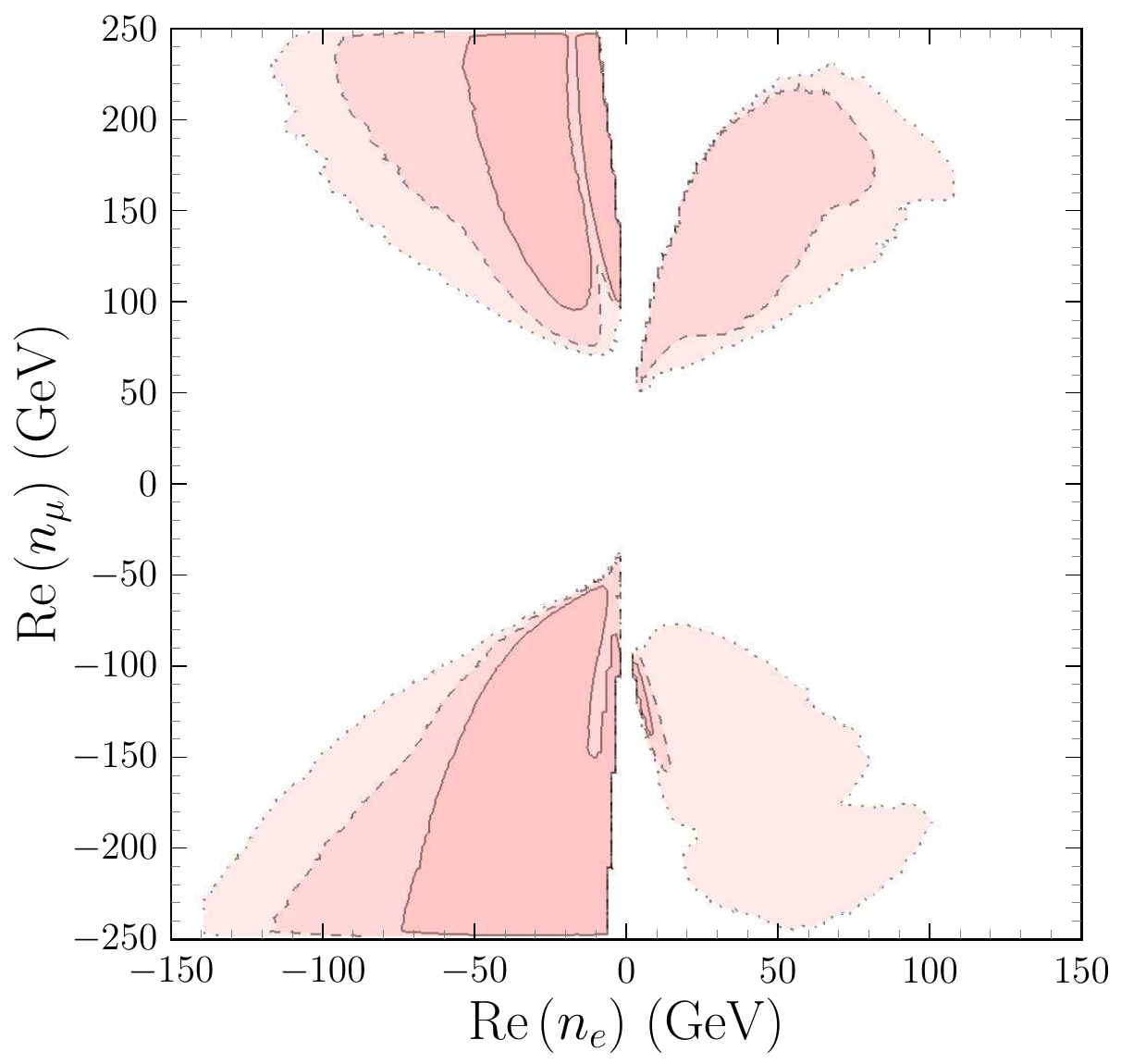}
\hfill
\includegraphics[width=0.32\textwidth]{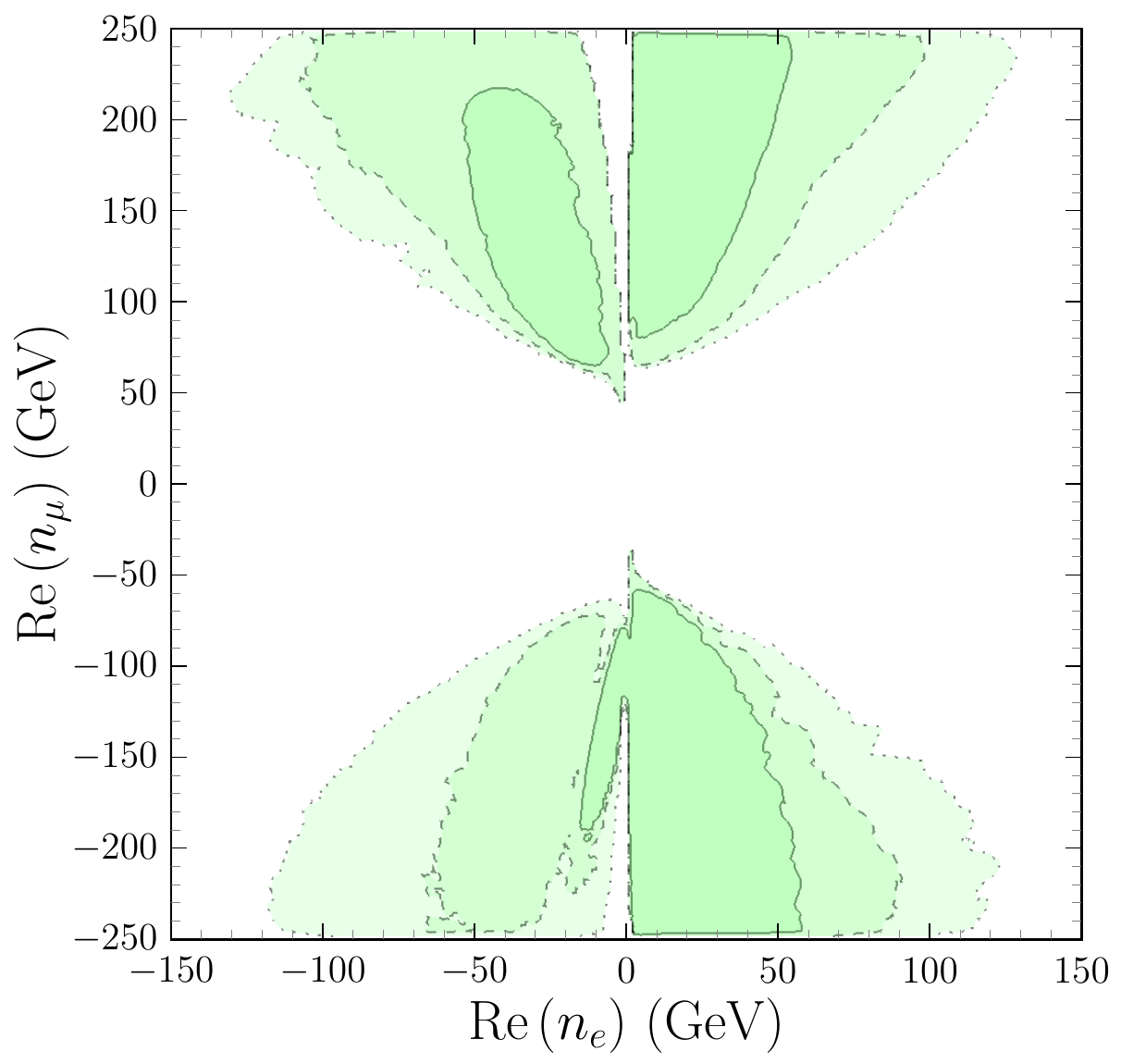}
\caption{Allowed regions of $\nrlm$ vs $\nrle$ for the tree studied scenarios concerning $\delta a_e^{\rm{Exp}}$. We follow the color scheme presented in Fig. \ref{Fig1}.\label{Fig:Scenarios}}
\end{center}
\end{figure}

A global fit to all the observables mentioned above is performed. To do so, a $\chi^2$-like function is sampled using Markov chain MonteCarlo techniques. The allowed regions of the parameter space are presented as 2D region plots where from darker to lighter the 1, 2 and 3$\sigma$ regions of $\Delta\chi^2=\chi^2-\chi^2_{Min}$ are shown. In Fig. \ref{Fig:Scenarios} the allowed regions for  $\nrlm$ vs $\nrle$ are presented for the three different scenarios studied. In blue, we use $\delta a_e^{\rm Exp,Cs}$ in \refeq{eq:dae:Cs}. In this case, the electron anomaly has the opposite sign to that of the muon anomaly. This usually entails a challenge for scalar mediators but in our case, we have a decoupling between electron and muon Yukawa couplings allowing us to explain both simultaneously. In red, we show the case with $\delta a_e^{\rm Exp,Rb}$ in \refeq{eq:dae:Rb}. Finally, in green, we present the situation where the electron anomaly is an average between the Cesium and Rubidium determinations. As we expected, the most constrained scenario is the analysis with $\delta a_e^{\rm Exp,Cs}$ and for that reason, we also find it the most attractive to study in-depth.

\begin{figure}[!htb]
\begin{center}
\includegraphics[width=0.32\textwidth]{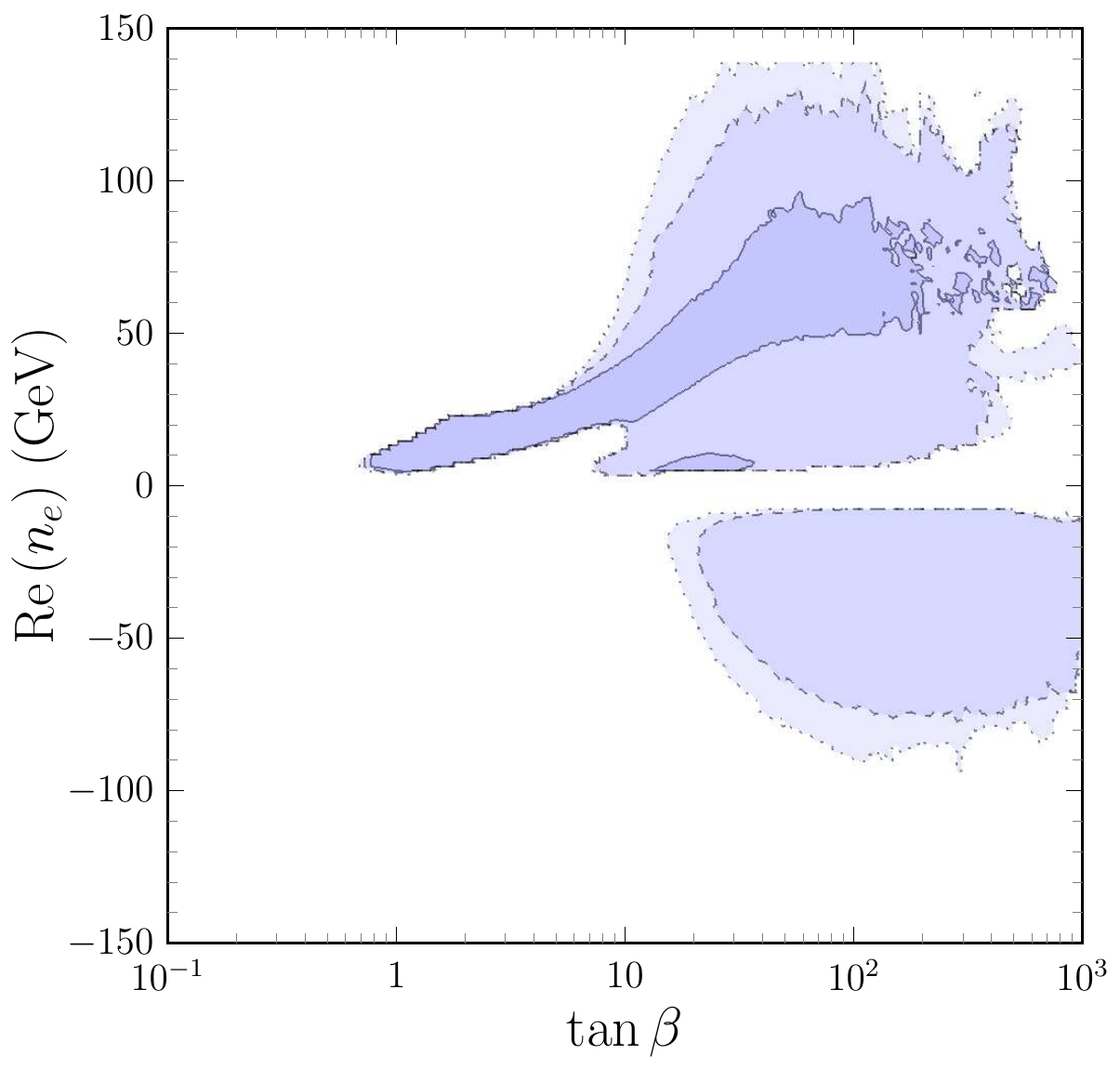}
\hfill
\includegraphics[width=0.32\textwidth]{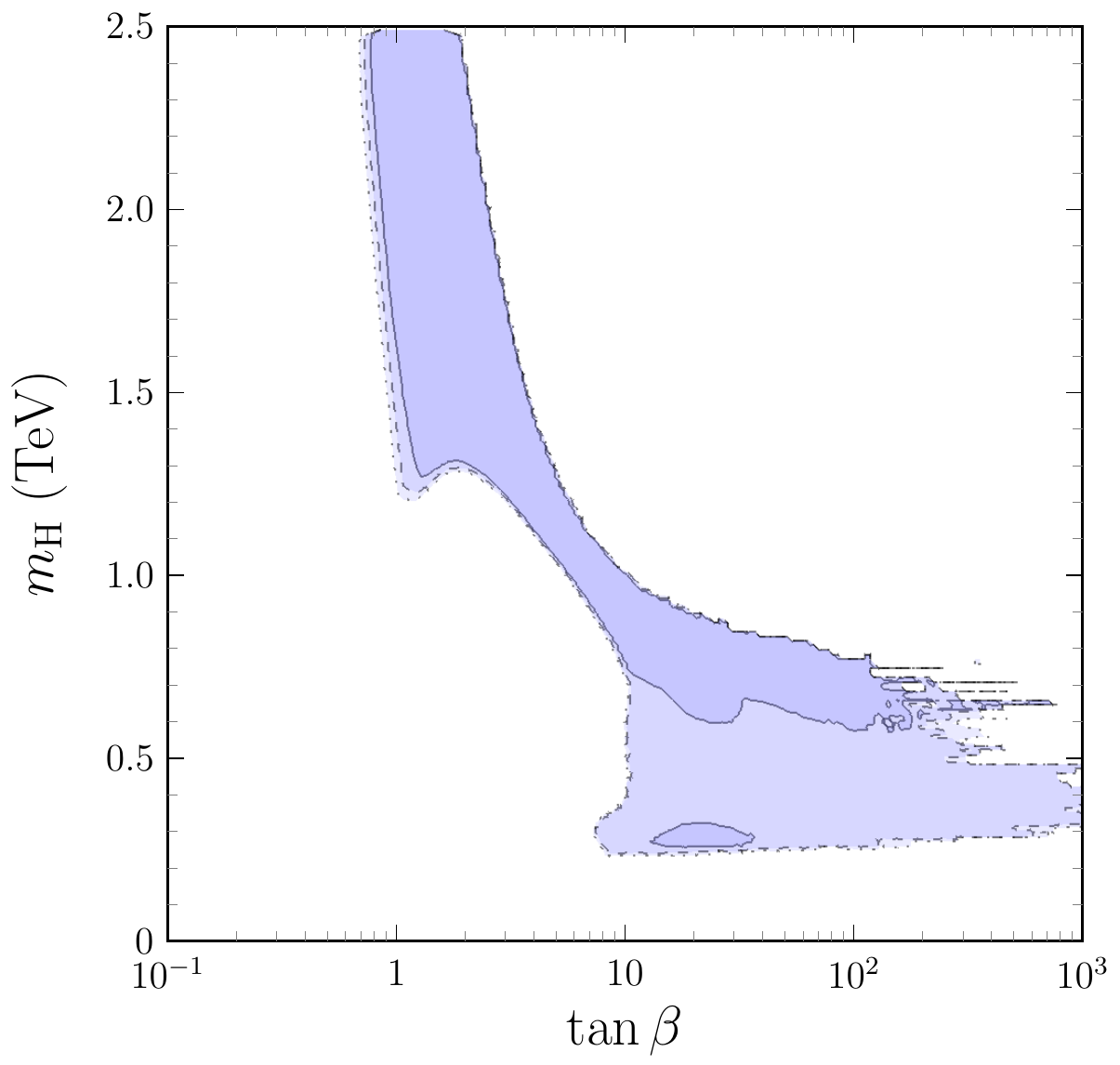}
\hfill
\includegraphics[width=0.32\textwidth]{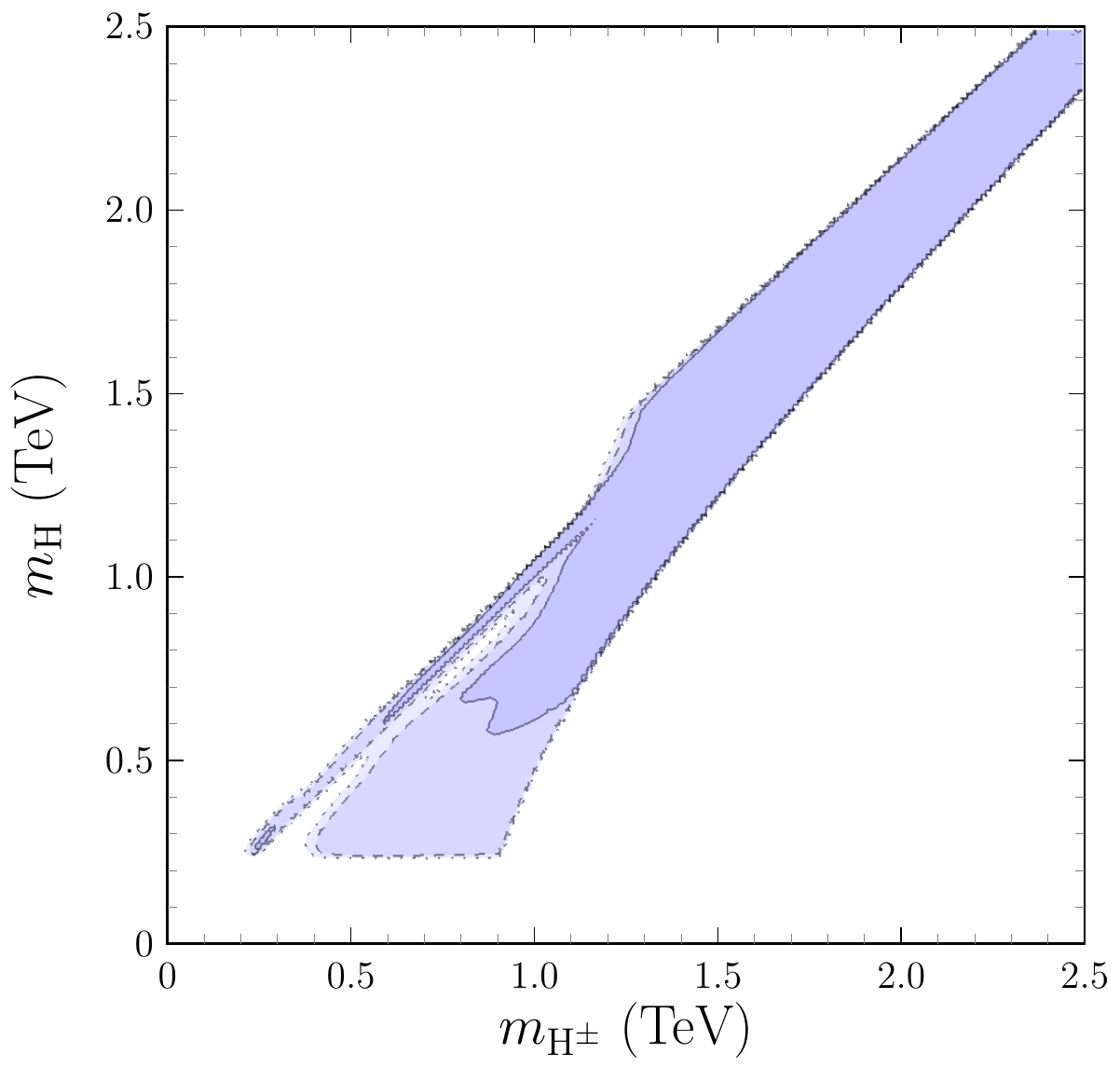}

\end{center}
\caption{Correlation among some of the parameters in the analysis with the the Cesium determination for the electron anomaly.\label{Fig:Plots:Cs}}
\end{figure}

Figure \ref{Fig:Plots:Cs} contains some relevant correlations between the couplings, $\tb$ and the scalar masses for the $\delta a_e^{\rm Exp,Cs}$ scenario. On the left, $\nrle$ vs $\tb$ is shown. We see that smaller values of $\tb$ require smaller couplings. It is also worth mentioning that for large $\tb$ the quark contribution to the $F$ function in  \refeq{eq:F:2loop} is very suppressed and then, the tau one becomes dominant. This allows us to have a negative sign for Re($n_e$) (if we have negative Re($n_\tau$)). On the center plot $\mH$ is presented as a function of $\tb$. We can see that the larger $\tb$ gets, the lighter the scalars are. On the right hand side, $\mH$ is plotted as a function of $\mcH$. We know that in this class of models, the oblique parameters constraint imposes that at low masses, there must be a degeneracy between $\mcH$ and $\mH$ or $\mA$. We can see in this figure that we get this below 1 TeV. We can see as well that in the case where $\cH$ must be degenerate with $\nA$ there is a lower bound of around 500 GeV. This prevents the electron anomaly to be explained at one-loop since \refeq{eq:deltal:1loop} forces to have $\nrle=\mA$ and this would violate perturbativity constraints ($\nrle\lesssim 250$ GeV). On the other hand, the muon one-loop solution would require $\nrlm=\mA/4$, and given the fact that there are available regions with $\mH<1$ TeV, perturbativity constraints do not suppose a problem in this case. We can identify then two scenarios, one where both anomalies are two-loop explained, in the heavy region, and a lighter region in mass where the electron remains two-loop but the muon is one-loop instead. 

\begin{figure}[!htb]
\begin{center}
\includegraphics[width=0.32\textwidth]{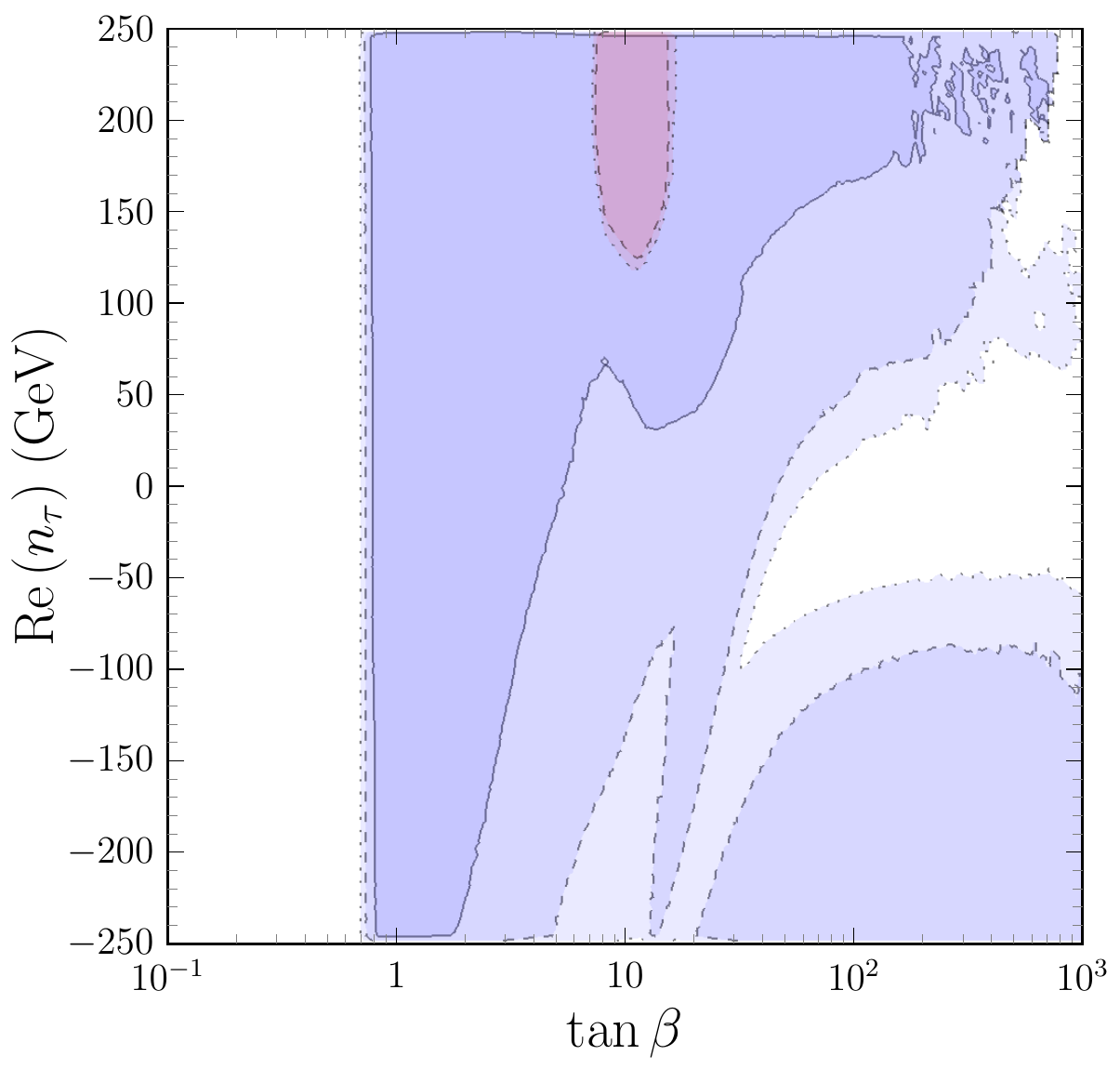}
\hfill
\includegraphics[width=0.32\textwidth]{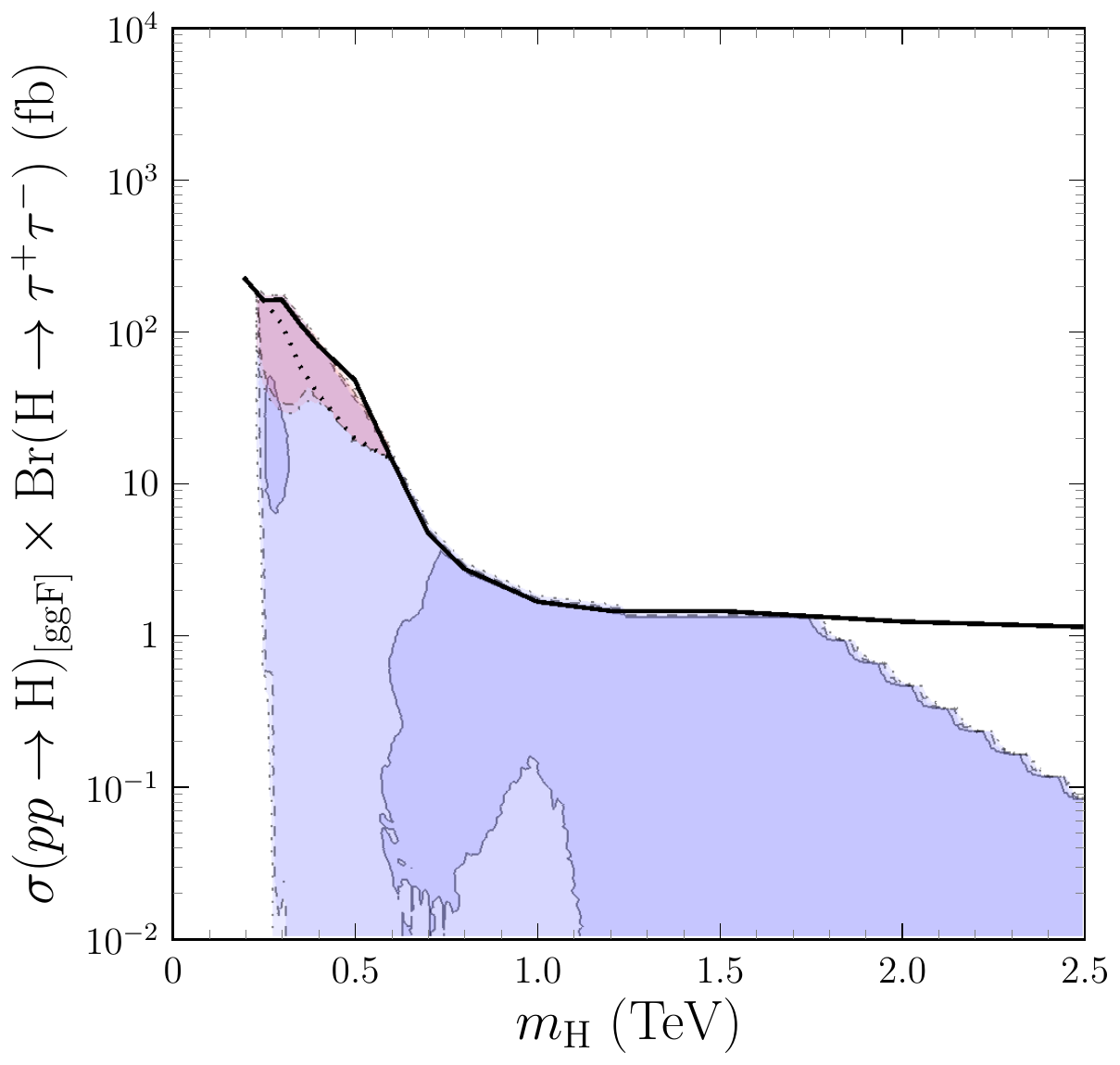}
\hfill
\includegraphics[width=0.32\textwidth]{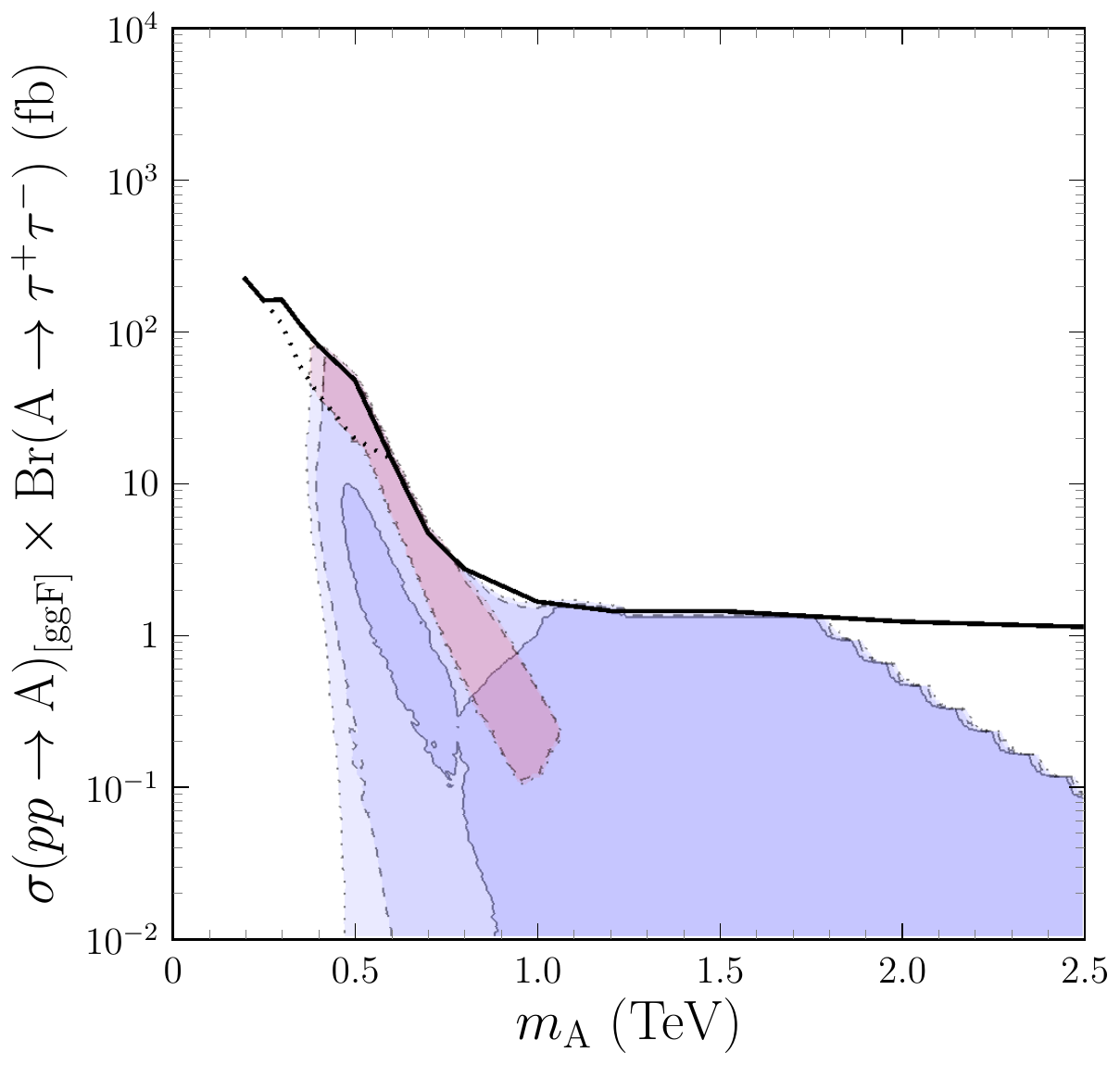}

\end{center}
\caption{Some correlations involving tau leptons. The regions where the excess in $\sigma(pp\to\nS)_{[\text{ggF}]}\times \text{Br}(\nS\to\tau^+\tau^-)$ is explained are plot in purple. \label{Fig:Plots:Tau}}
\end{figure}

Figure \ref{Fig:Plots:Tau} contains some relevant information regarding the tau lepton. In the left plot, $\nrlt$ is presented versus $\tb$. As we can see, the tau coupling is almost unconstrained, even more in the low $\tb$ region. For that reason, we asked ourselves if there was another anomaly regarding taus that this model could give an explanation to. In purple, we show the region of the parameter space that is in agreement with the observed excess by ATLAS in the direct search of a heavy Higgs ($\mS{} \sim 400$ GeV) in the di-tau channel \cite{ATLAS:2020zms}. In the center and right figures we can see the gluon-gluon fusion (ggF) production cross section of a scalar $\nH$ (center) and a pseudoscalar $\nA$ (right), times the branching ratio to $\tau^+\tau^-$ with respect to the mediator mass. The black dotted line corresponds to the expected limit while the solid one is the observed limit by ATLAS. The purple regions of Fig. \ref{Fig:Plots:Tau} correspond to scenarios where this excess is explained with $\nH$ or $\nA$ in the model presented here.

\section{Summary and conclusions}

We have presented a flavor conserving two Higgs doublet model that shares the quark sector with a type I or X but that is general flavor conserving in the lepton sector. This model has been proved to be one-loop stable under renormalization. This framework is capable of accommodating both the electron and muon ($g-2$) anomalies, including the different determinations of the electron one. In the region where the scalars are heavier than 1 TeV, the two-loop contributions for both observables are dominating. On the other hand, for lighter scalar and $\tb\gg 1$, the muon anomaly is one-loop dominated while the electron remains two-loop due to perturbativity and universality constraints. This model is also capable of explaining the small excess observed by ATLAS in $\sigma(pp\to\nS)_{[\text{ggF}]}\times \text{Br}(\nS\to\tau^+\tau^-)$ at around 400 GeV.

\section*{Acknowledgments}

We want to thank the organizers of the 56th Rencontres the Moriond for setting up the conference and for giving FCG the opportunity of sharing our work. The authors acknowledge support from Spanish \textit{Agencia Estatal de Investigación--Ministerio de Ciencia e Innovación} (AEI-MICINN) under grants PID2019-106448GBC33 and PID2020-113334GB-I00/AEI/10.13039/501100011033 (AEI/FEDER, UE) and
from \textit{Generalitat Valenciana} under grant PROMETEO 2019-113. FCG acknowledges the \textit{MICINN}, Spain (Grant BES-2017-080070) for supporting his work. CM is supported by \textit{Conselleria de Innovación, Universidades, Ciencia y Sociedad Digital} from \textit{Generalitat Valenciana}
(grant ACIF/2021/284). MN is funded by the GenT Plan from \textit{Generalitat Valenciana} under project CIDEGENT/2019/024.

\end{document}